\documentclass[12pt,preprint]{aastex}
\usepackage{lscape}

\shorttitle{ANISOTROPIC RADIATIVE PRESSURE FEEDBACK BY AGNS}

\shortauthors{LIU \& ZHANG}

\begin{document}


\title{Dusty torus formation by anisotropic radiative pressure feedback of active galactic nuclei}


\author{Yuan Liu and Shuang Nan
Zhang}

\affil{Key Laboratory of Particle Astrophysics, Institute of High
Energy Physics, Chinese Academy of Sciences, P.O.Box 918-3, Beijing
100049, China}

\email{liuyuan@ihep.ac.cn; zhangsn@ihep.ac.cn}


\begin{abstract}
The feedback by active galactic nuclei (AGNs) is significant for the
formation and evolution of galaxies. It has been realized that the
radiative pressure feedback could be an efficient mechanism due to
the existence of dust. In this Letter, we discuss the effect of
anisotropic radiative pressure, which is inevitable if the
UV/optical emission arises from an accretion disk. The distribution
of dusty gas should be also anisotropic due to the influence of the
anisotropic disk radiation, i.e. the dust in the face-on direction
of an accretion disk can be blown out relatively more easily,
whereas the dust can survive in the edge-on direction. This result
can explain the presence of some obscured
 AGNs with high Eddington ratios and
can also quantitatively reproduce the observed decreasing fraction of type 2 AGNs with increasing luminosity. A
sequence of AGN formation and evolution is also proposed, within the context of the formation, evolution and
exhaustion of the dusty torus. Our model predicts the existence of bright AGNs with dusty tori, but without
broad line regions. Finally we discuss the implications of the anisotropic radiation for the calculations of
luminosity functions and radiation efficiencies of AGNs.
\end{abstract}



\section{Introduction}\label{int}
Various correlations over the past decade have been found between
the mass of black holes  in the center of  galaxies ($M_{\rm{BH}}$)
and the properties of host galaxies, e.g. the velocity dispersion
$\sigma$ of the galaxy's bulge ($M_{\rm{BH}}$-$\sigma$ relation,
Tremaine et al. 2002), the mass of  the galaxy's bulge
$M_{\rm{bulge}}$
 ($M_{\rm{BH}}$-$M_{\rm{bulge}}$ relation,
 Marconi \& Hunt 2003), and the concentration  of the spheroid (Graham \& Driver
2007). These results indicate that supermassive black holes play a
fundamental role in the formation and evolution of galaxies (Silk \&
Rees 1998; Fabian et al. 2002; Di Matteo et al. 2005; Menci et al.
2008). Numerous models have been proposed to explain the observed
correlations. The  feedback by active galactic nuclei (AGNs) is
promising to connect the properties at quite different scales (e.g.
King 2003; Murray et al. 2005; Springel et al. 2005; Shin et al.
2010).

The form of feedback could be either  energy injection or momentum
injection. Among various mechanisms of feedback, radiative pressure
could be an efficient one, though it is only important when the
luminosity of the central source is near the Eddington luminosity,
i.e. $L_{\rm{Edd}}=4\pi G m_{\rm{p}} c M_{\rm{BH}}/\sigma_{\rm{T}}$,
where $G$ is the gravitational constant, $m_{\rm{p}}$ is the proton
mass, $c$ is the light speed, and $\sigma_{\rm{T}}$ is the
cross-section for Thomson scattering. However, the existence of dust
could remarkably amplify the effective cross-section and then reduce
the required luminosity that can balance the gravitational force of
the central black hole. If the effective cross-section of dusty gas
is $\sigma_{\rm{eff}}=A\sigma_{\rm{T}}$, where $A>1$ is the boost
factor, the effective Eddington luminosity is simply lower than
$L_{\rm{Edd}}$ by a factor of $A$. Thus the feedback by radiative
pressure can be important even if the real luminosity is
sub-Eddington. Therefore, the dusty gas will be cleared out by the
accreting black hole in a very short time ($\lesssim10^5$ yr) if the
luminosity of the accreting black hole is high enough (Chang et al.
1987; Murray et al. 2005; Fabian et al. 2002, 2006; Raimundo et al.
2010).

Although the importance of radiative pressure has been discussed by several authors, the effect of anisotropic
radiation has so far not been taken into account. As we will show in this Letter, including the effect of
anisotropic radiative pressure is critical to understanding several observational results. Actually, anisotropic
radiation is a natural result if the emitting region is an accretion disk, especially for the ultraviolet and
optical emission, which are thought as the evident signal of an accretion disk around a supermassive black hole.

In \S\ref{dust}, we investigate the effects of anisotropic radiative pressure on the distribution of dust. In
\S\ref{int}, we utilize the anisotropic radiative pressure to explain the presence of some AGNs with high
Eddington ratios and high column densities. As a further quantitative test, we reproduce the observed fraction
of type 2 AGNs using the model proposed in this Letter  in \S\ref{type2}. In \S\ref{con}, we discuss the
implications of the anisotropic radiation and give our conclusions.

\section{Anisotropic radiative pressure}\label{dust}
The big blue bump in ultraviolet to optical bands of the spectral
energy distribution of AGNs dominates the total output of AGNs and
is thought as the emission from an accretion disk around a
supermassive black hole, though the observational evidence is still
not very unambiguous (e.g. Shang et al. 2005; Kishimoto et al.
2008). The UV/optical opacity also dominates the overall opacity in
the dust. As a result, the anisotropic disk UV/optical emission must
have significant impacts on the global properties of the dusty
structure around the accretion disk. For the standard accretion disk
model, which is optically thick and geometrically thin, the optical
depth is dominated by electron scattering in the inner part, which
is the emitting region of ultraviolet and optical photons (Shakura
\& Sunyaev 1973). In this case, the emitting specific intensity $I$
depends on the inclination angle $\mu=\cos\theta$, where $\theta$ is
the angle between the line of sight and the normal of the accretion
disk, as $I(\mu)\propto(1+2\mu)$ (Chandrasekhar 1960; Sunyaev \&
Titarchuk 1985). Therefore, the observed flux is
\begin{equation}\label{flux}
   F(\mu ) = \frac{L}{{4\pi r^2 }}\frac{{\mu I(\mu )}}{{\int_0^1
   {\mu I(\mu )d\mu } }} = \frac{6}{7}F_0 \mu (1 + 2\mu )
,
\end{equation}
where $L$ is the total luminosity of the source, $r$ is the distance
between the observer and the source, and $F_0=L/4\pi r^2$. Equation
(\ref{flux}) is valid only if the optical depth is large enough,
i.e. $\tau>10$, which is well fulfilled in the inner region of the
standard accretion disk model. In contrast to the anisotropic disk
UV/optical continuum, the emitting region of X-ray is likely to be
optically thin or a quasi-spherical corona, hence the X-ray emission
should be nearly isotropic.

The relativistic effects can diminish the anisotropy of the photons from the innermost region of the accretion
disk around a rapidly spinning black hole; for example, the edge-on flux can be effectively enhanced by the
gravitational bending of the black hole. However, these effects are sensitive to the frequency of photons and
the value of the spin of black holes in AGNs is still under debate (Volonteri et al. 2005; Wang et al. 2009;
Shankar et al. 2009). In contrast, the flaring structure in the outer part of the accretion disk can shield the
radiation from the inner part (Shakura \& Sunyaev 1973), and therefore may partly cancel the gravitational
bending effect and then enhance the anisotropy. Thus we temporarily ignore the above effects in this Letter for
simplicity.

Since the radiation flux depends on the inclination angle, the effective Eddington luminosity should also be a
function of the inclination angle. The equation of radial motion for the material under the influence of the
anisotropic radiative pressure and the gravitational force of black hole is
\begin{equation}\label{lumi}
\frac{{dv}}{{dt}} = \frac{{\sigma _{{\rm{eff}}} F(\mu
)}}{{m_{\rm{p}} c}} - \frac{{GM_{{\rm{BH}}} }}{{r^2 }}.
\end{equation}
When the gravitational force is balanced by the radiative pressure,
we obtain an equation of the critical angle $\theta_c$
\begin{equation}\label{ang}
\cos \theta _{\rm{c}}  = \frac{{\sqrt {1 + 28 /(3\lambda)}  -
1}}{4},
\end{equation}
where the effective Eddington ratio $\lambda  = AL/L_{{\rm{Edd}}}$.

\begin{figure}
\begin{center}
\includegraphics[scale=0.5]{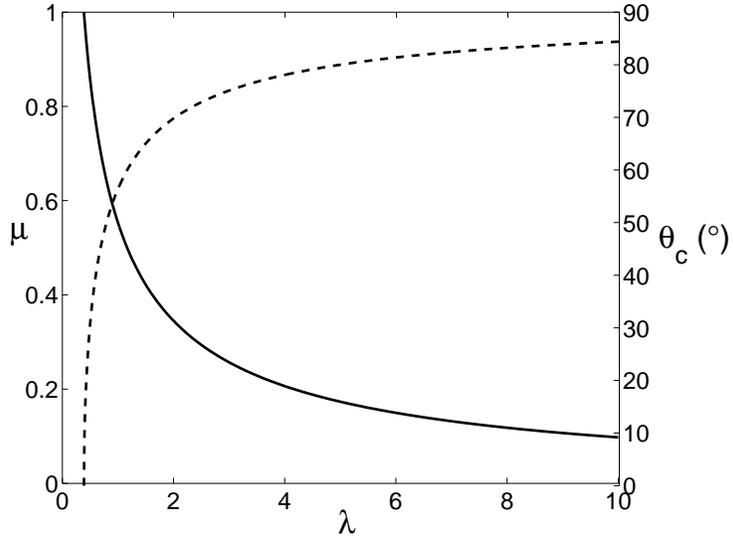}
\caption{\label{fig1} The critical angle $\mu=\cos\theta_{\rm{c}}$
as a function of $\lambda=AL/L_{\rm{Edd}}$.}
\end{center}
\end{figure}

\begin{figure}
\begin{center}
\includegraphics[scale=0.5]{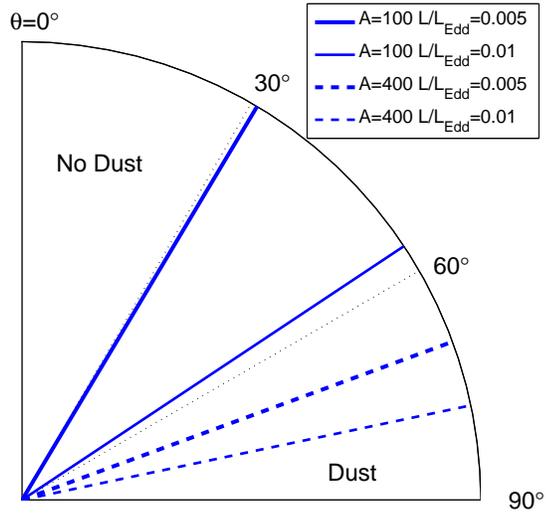}
\caption{\label{fig2} The section plan of the distribution of dust
with different values of $A$ and $L/L_{\rm{Edd}}$ in
spherical coordinate. The lines are dividing planes with $\theta=\theta_c$.}
\end{center}
\end{figure}

We show the dependence of $\theta_c$ on $\lambda$ in Figure
\ref{fig1} and the distribution of dust for different values of $A$
and $L/L_{\rm{Edd}}$ in Figure \ref{fig2}. If $\lambda<7/18$, there
is no real root for equation (\ref{ang}), i.e., the radiative
pressure is too weak to blow out the material in any direction. If
$\lambda\geq7/18$, then the matter between $0$ and $\theta_c$ will
be expelled by the radiative pressure. However, the matter
distributed with $\theta>\theta_c$ will still be controlled by the
gravitational force. We note that this critical value of $\lambda$
is smaller than unity, which is the result of the anisotropic
radiation, i.e., more energy is emitted towards $\mu=1$. At the same
time, even the luminosity is very high ($\lambda\sim10$), there is
still a considerable solid angle ($\sim$0.1) within which the dust
could survive. In general, as the result of the anisotropic
radiative pressure feedback, the distribution of dust is also
anisotropic, i.e., the dust is expelled in the face-on direction of
the accretion disk but still exists in the edge-on direction. Since
gas is strongly coupled with dust by Coulomb force, scattering, and
magnetic field (Chang et al. 1987; Scoville \& Norman 1995; Murray
et al. 2005), we expect the distribution of gas should be similar to
that of dust.

As we will show in the next two sections, this anisotropic distribution of dusty gas is important for explaining
several observational results.

\section{Column density vs. Eddington ratio plane}\label{int}
In the isotropic radiative pressure model, it is predicted that there is a dividing line in the column density
($N_{\rm H}$) vs. Eddington ratio ($L/L_{\rm{Edd}}$) plane; at the right side of the plane, i.e. the Eddington
ratio is high enough, the dusty gas could not be long-lived (e.g. see Figure 4 in Raimundo et al. 2010).
However, the presence of obscured quasars is likely to contaminate the right side of the dividing line in the
$N_{\rm H}$-$L/L_{\rm{Edd}}$ plane (e.g. Brusa et al. 2005; Vignali et al. 2006; Polletta et al. 2008). Although
the contamination may be explained as the absorption in the large scale of host galaxies, outflow or transient
absorption (Raimundo et al. 2010), we propose that this discrepancy could be more naturally alleviated by the
effect of anisotropic radiative pressure. As discussed in \S\ref{dust}, even if the Eddington ratio is high, in
the edge-on direction of the accretion disk, there still exists a region dominated by the gravitational force
and therefore the dust could exist. If our line of sight happens to intersect this region, we will observe an
obscured AGN. Therefore, there is no definite dividing line in the $N_{\rm H}$-$L/L_{\rm{Edd}}$ plane. However,
the probability of the presence of such a high column density AGN decreases as the increase of Eddington ratio
and thus the decrease of the solid angle occupied by dust (see Figure 1 and 2).

\begin{figure}
\begin{center}

\includegraphics[scale=0.5]{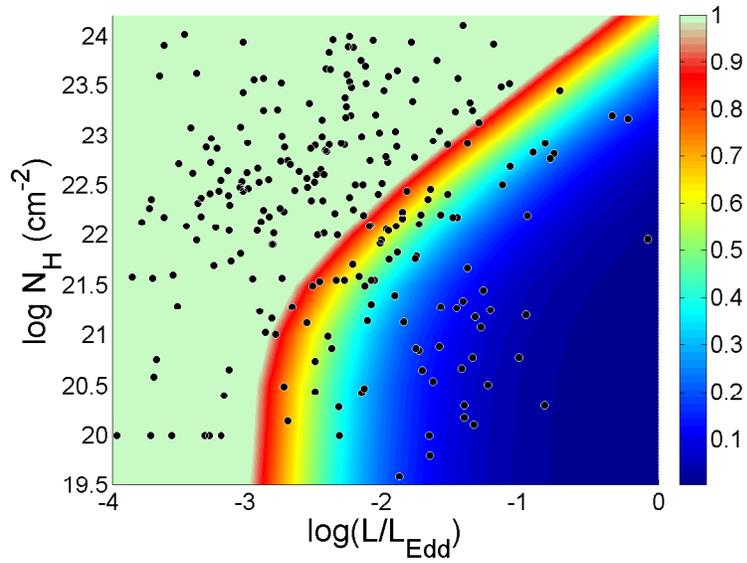}
\caption{\label{fig3} The probability, shown in color scale, of
finding an obscured AGN in the $N_{\rm H}-L/L_{\rm{Edd}}$ plane
considering the anisotropic radiative pressure. The data points
(\textit{black} circles) mark locations of AGNs on the $N_{\rm
H}-L/L_{\rm{Edd}}$ plane, taken from Figure 4 in Raimundo et al.
(2010) (the data points only with the upper limits of $N_{\rm H}$
are ignored).}
\end{center}
\end{figure}

Figure \ref{fig3}  shows the probability of finding an obscured AGN in the $N_{\rm H}$-$L/L_{\rm{Edd}}$ plane.
We adopt the relation between $N_{\rm H}$ and $A$ obtained by Fabian et al. (2006). Since the value of the boost
factor $A$ is smaller for higher column density, it is more likely to find an obscured AGN with high
$L/L_{\rm{Edd}}$ in the upper part of  the $N_{\rm H}$-$L/L_{\rm{Edd}}$ plane. Although the probability
decreases in the right part of  the $N_{\rm H}$-$L/L_{\rm{Edd}}$ plane, there is still a considerable
probability (at least 0.1$\sim$0.2) to find AGNs with high $N_{\rm H}$ in the forbidden region obtained by
assuming isotropic radiation from an accreting black hole located at the center of an AGN.

This prediction is more consistent with the observed result shown in
 Figure \ref{fig3}. Note that the bolometric luminosity of the observed sample is estimated by
 the more isotropic X-ray
 luminosity and thus can be considered as the intrinsic one, or at least tracks the intrinsic one well
 (as we discussed in \S\ref{dust}).
  Although the uncertainty of $L/L_{\rm{Edd}}$ is unlikely
  to qualitatively change the distribution of the observed sample, it is
  not
straightforward to compare our prediction quantitatively in the
$N_{\rm H}-L/L_{\rm{Edd}}$ plane due to the unknown underlying bias
and intrinsic distributions of $L/L_{\rm{Edd}}$ and $N_{\rm H}$.
Another closely related observed result, the decreasing fraction of
type 2 AGNs with increasing luminosity, provides a good opportunity
to test our scenario quantitatively. We will address this problem in
the next section.

\section{The fraction of Type 2 AGNs }\label{type2}
In the unified model of AGNs, the classification is determined
solely by the viewing angle between the observer and the symmetry
axis of the dusty torus (Antonucci 1993). For type 1 AGNs, the broad
line region (BLR) and the central engine are observed directly
through the opening angle of the dusty torus, but they are obscured
by the torus for type 2 AGNs. Thus, it is expected that the
intrinsic properties of both type 1 and 2 AGNs are the same.
However, with the fast growth of multi-wavelength surveys, it is
found that the fraction of type 2 AGNs decreases with increasing
luminosity (e.g, Ueda et al. 2003; Maiolino et al. 2007; Hasinger
2008). We propose here that the observed fraction of type 2 AGNs
could be explained if we consider the effects of the anisotropic
 UV/optical radiation from the accretion disk. As we discussed in
\S\ref{dust}, dust could only exist in the region with
$\theta_c<\theta<\pi/2$ and the covering factor of this region will
decrease as the luminosity of the central source increases for the
given black hole mass.

We adopt the fraction of type 2 AGNs obtained by Hasinger (2008),
who compiled the largest AGNs sample (1290 AGNs) from several X-ray
surveys (2-10 keV), combined the optical and X-ray classifications,
and then gave the most accurate estimate of the fraction of type 2
AGNs up to now.

For a realistic sample of AGNs, the observed luminosity is also correlated with the black hole mass. We simply
parameterize this relation as $M \propto L^\alpha$; we then combine the normalization of this relation, the
booster factor $A$, and the conversion factor between $L$ and $L_{\rm{X}}$ (we assume $L \propto L_{\rm{X}}$,
where $L_{\rm{X}}$ is the X-ray luminosity in 2-10 keV band) into a free parameter, the value of which is
adjusted to match the observation data. If we identify the fraction of the solid angle covered by the dust as
the fraction of type 2 AGNs ($f_2$), then $f_2=\cos\theta_{\rm{c}}$ and $f_2$ is a function of $L_{\rm{X}}$
according to equation (\ref{ang})

\begin{figure}
\begin{center}
\includegraphics[scale=0.5]{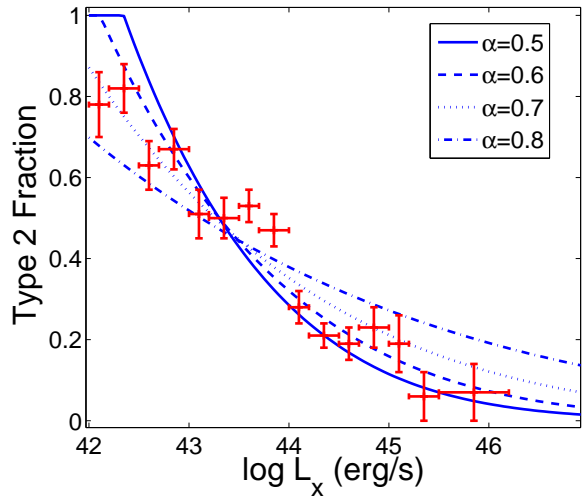}
\caption{\label{fig4} The fraction of type 2 AGNs as a function of
X-ray luminosity $L_{\rm{x}}$ in 2-10 keV band. The data points with
errors bars are taken from Hasinger (2008). }
\end{center}
\end{figure}

The real value of $\alpha$ of this sample is unclear. Thus we try several values of $\alpha$ in Figure
\ref{fig4}, but do not treat it as a free parameter since we do not expect a single value of $\alpha$ could
account for all AGNs in the sample. It seems that $\alpha\sim0.6-0.7$ can well reproduce the observed fraction
and a larger value of $\alpha$ is more appropriate in the low luminosity band.  Actually, the fraction of type 2
AGNs could be overestimated in our model, especially in the low luminosity band, since the collapse of dust is
not considered (see the discussion in \S\ref{con}). The value of $\alpha$ also depends on samples, which may
explain the somewhat different trends of type 2 AGNs found in different samples (e.g. see the comparison in
Hasinger 2008).
%%\subsection{Observations}\label{sec21}

\section{Discussion and conclusion}\label{con}
\begin{figure*}
\begin{center}
\includegraphics[width=\textwidth,keepaspectratio]{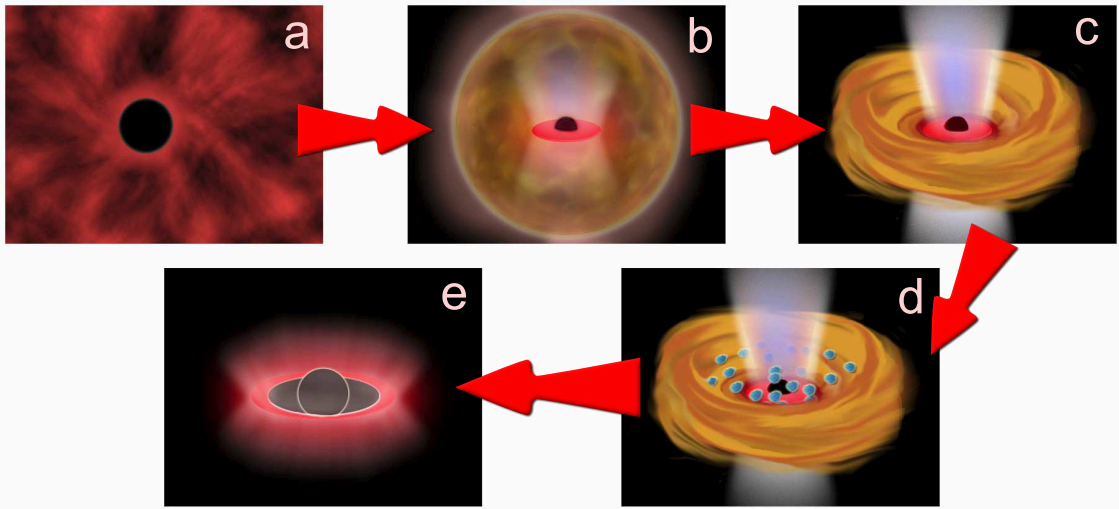}
\caption{\label{fig5} Illustration of the formation and evolution of
AGNs. (\textit{a}) The fast growth phase of a seed black hole.
(\textit{b}) The black hole and its accretion disk are embedded in a
dusty ball. (\textit{c}) The anisotropic radiative pressure blows
out the dusty gas in the face-on direction of the accretion disk and
results in a dusty torus. (\textit{d}) The normal phase of an AGN
with the inner structure, e.g.  broad line region. (\textit{e}) The
dying phase of an AGN when its fuel is consumed out. See text for
further explanations.}
\end{center}
\end{figure*}

In this Letter, we have investigated the influence of the
anisotropic radiation from the accretion disk on the distribution of
dust in AGNs. The critical luminosity required to blow out the dusty
gas is smaller than the standard Eddington luminosity (including the
effects of dust), since more radiation is concentrated towards the
face-on direction of the accretion disk. Therefore, there is a
critical angle defining a permitted region where the dust could
exist even the Eddington ratio is high. Thus, we obtained the
probability of finding an obscured AGN in the $N_{\rm
H}$-$L/L_{\rm{Edd}}$ plane, which is quantitatively consistent with
observations. We further tested our model using the fraction of type
2 AGNs determined by X-ray surveys and found the decreasing trend of
the fraction with increasing luminosity can be well explained.

Under our scenario, there should be an evolving sequence from type 2
AGNs to type 1 AGNs (Koulouridis et al. 2009), as illustrated in
Figure~5. It is commonly believed that seed black holes must be
formed before the AGN phase (Figure~5~(a)) (e.g. Lu et al. 2003; Hu
et al. 2006). In the initial AGN phase, the dust produced by strong
star formation process could shield the emission from AGNs from
almost all directions (Figure~5~(b)); therefore in this phase most
AGNs appear as type 2s. Subsequently, the feedback from AGNs begin
to expel the dusty gas (Figure~5~(c)) and then the type 1 AGNs begin
to dominate the bright population of AGNs, as shown in Figure
\ref{fig4}. Therefore, there should be a significant increase in the
fraction of type 2 AGNs with redshift, which is supported by recent
observations (Hasinger 2008; Treister et al. 2010). As gas is
strongly coupled with dust, both dust and gas should be blown out
within the radiation cone. At this stage the AGN should have a dusty
torus, but no BLR (Figure~5~(c)). Our model describes the physical
process from stage (b) to stage (c) in Figure~5, and thus {\it
predicts the existence of bright AGNs with dusty tori, but without
BLRs},  e.g. weak line quasar (Fan et al. 1999; Plotkin et al.
2010). Subsequently the BLR is formed, as the dusty torus supplies
the fuel to the accretion disk (Figure~5~(d)). As the fuel from the
torus is gradually exhausted, both the torus and the BLR disappear
at the end of the AGN activity (Figure~5~(e)), in qualitatively
agreement with the results by Nicastro (2000), Laor (2003), Elitzur
and Ho (2009), Zhang et al. (2009), and Zhu et al (2009); we will
investigate quantitatively the effects of anisotropic radiative
pressure on this evolutional scenario in a future work.

If the dusty torus could exist for a long time, there should be some supporting mechanism to maintain its
thickness. However, this is still a controversial issue. Several models are proposed, e.g., random motion of
clouds, radiative pressure, and magnetic field (see Krolik 2007 and references therein). Strictly speaking, the
permitted region given by equation (\ref{ang}) is an upper limit for the height of the dusty torus. The real
height of the torus could be smaller than this region if the support is not sufficient to make the dust fill
this region, e.g., the disk wind may quench in low luminosity AGNs (Elitzur \& Shlosman 2006). However, as shown
in \S\ref{type2}, the observed fraction of type 2 AGNs could be well explained by this simple model, which
indicates that the support is likely to be sufficient in most cases. We must stress that this statement is only
tentative. More realistic model of the torus is beyond the scope of this Letter and will be discussed in future
works.

The covering factor of the dusty torus (or the fraction of type 2 AGNs) can  be also inferred from the ratio
between mid-IR to bolometric luminosity. Maiolino et al. (2007) and Treister et al. (2008) found that the
covering factors determined by this method and X-ray surveys have the same trends with the bolometric
luminosity, but the former is systematically higher. If the above result is real, this could be due to the
missing Compton thick AGNs in X-ray surveys and well explained by our model, since the value of the boost factor
$A$ is smaller for Compton thick AGNs (large $N_{\rm H}$). However, the contribution from other components
rather than the dusty torus can contaminate the mid-IR luminosity and further make the covering factor estimated
in this way questionable (Rowan-Robinson et al. 2009; Mor et al. 2009). The inconsistency in different
observations may also reveal the complex geometry of material responsible for absorptions in X-ray and mid-IR
bands, which should be further investigated using multi-wavelength data.

Our model is not in conflict with the receding torus model but
actually complementary to that model (Lawrence 1991). Due to the
unclear supporting mechanism and geometry of dust, the fraction of
type 2 AGNs cannot be directly related to the variation of the inner
radius of dust. The radiative pressure, as another constraint on the
permitted region of dust, could link the opening angle of the dust
(the fraction of type 2  AGNs) to the properties of the black hole
more explicitly. Our model also predicts that  the direction of the
symmetry axis of the dust torus will be consistent with that of the
normal of the accretion disk, but may be random relative to the
direction of the host galaxy.

%According to the result in \S\ref{dust}, when the Eddington ratio is
%large enough, the covering factor of torus will be very small. For
%the standard accretion disk model, $H/R\propto \alpha^{-1/10}
%R^{1/8}$ where $H$ and $R$ are the half-thickness and radius of the
%disk respectively and $\alpha$ is the viscosity parameter (Shakura
%\& Sunyaev 1973). Thus, the outer part of the disk can shield the
%radiation from the inner part and form a considerable permitted
%region of dust if $\alpha\ll1$ in the neutral disk (Gammie 1996).

The anisotropic radiation could be important for the estimate of the intrinsic luminosity and then several
properties of AGNs. For example, for an AGN observed with an inclination angle $\theta=30^\circ$, by equation
(\ref{flux}), the intrinsic luminosity is overestimated by a factor of two, if we calculate the luminosity using
the observed flux assuming isotropic radiation. We will consider the effect of this bias on the calculations of
luminosity function, radiation efficiency, and further black hole spin in future works.

 \acknowledgments{}
 We thank Sandra Raimundo for providing the data of column density and Eddington ratio used in
 Figure 3.
  S.N.Z. thanks Ari Laor for initial discussions on this idea, encouragement for pursuing this work, and commenting on the draft manuscript.
We also appreciate comments from the anonymous referee. S.N.Z. acknowledges partial funding support the
Directional Research Project of the Chinese Academy of Sciences under project no. KJCX2-YW-T03 and by the
National Natural Science Foundation of China under grant nos. 10821061, 10733010, 10725313, and by 973 Program
of China under grant 2009CB824800.


\begin{thebibliography}{}
\bibitem[]{} Antonucci, R. R. 1993, ARA\&A, 31, 473
\bibitem[]{} Brusa, M., et al. 2005, A\&A, 432, 69
\bibitem[]{} Chandrasekhar, S. 1960, Radiative Transfer (New York: Dover)
\bibitem[]{} Chang, C. A., Schiano, A. V. R., \& Wolfe, A. M. 1987, ApJ, 322, 180
\bibitem[]{} Di Matteo, T., Springel, V., \& Hernquist, L. 2005, Nature, 433, 604
\bibitem[]{} Elitzur, M., \& Shlosman, I. 2006, ApJ, 648, L101
\bibitem[Elitzur \& Ho(2009)]{2009ApJ...701L..91E} Elitzur, M., \& Ho, L.~C.\ 2009, \apjl, 701, L91
\bibitem[]{} Fabian, A. C., Wilman, R. J., \& Crawford, C. S. 2002, MNRAS, 329, L18
\bibitem[]{} Fabian, A. C., Celotti, A., \& Erlund, M. C. 2006, MNRAS, 373, L16
\bibitem[]{} Fan, X., et al. 1999, ApJ, 526, L57
\bibitem[]{} Graham, A. W., \& Driver, S. P. 2007, ApJ, 655, 77
\bibitem[]{} Hasinger, G. 2008, A\&A, 490, 905
\bibitem[Hu et al.(2006)]{2006MNRAS.365..345H} Hu, J., Shen, Y., Lou, Y.-Q., \& Zhang, S.\ 2006, \mnras, 365, 345
\bibitem[]{} King, A. 2003, ApJ, 596, L27
\bibitem[]{} Kishimoto, M., Antonucci, R., Blaes, O., Lawrence, A., Boisson, C., Albrecht, M., \& Leipski, C. 2008, Nature, 454, 492
\bibitem[]{} Koulouridis, E., Plionis, M., Chavushyan, V., Dultzin, D., Krongold, Y., Georgantopoulos, I., \& Goudis, C. 2009, arXiv:0910.1355
\bibitem[]{} Krolik, J. H. 2007, ApJ, 661, 52
\bibitem[Laor(2003)]{2003ApJ...590...86L} Laor, A.\ 2003, \apj, 590, 86
\bibitem[]{} Lawrence, A. 1991, MNRAS, 252, 586
\bibitem[Lu et al.(2003)]{2003ApJ...590...52L} Lu, Y., Cheng, K.~S., \& Zhang, S.~N.\ 2003, \apj, 590, 52
\bibitem[]{} Maiolino, R., Shemmer, O., \& Imanishi, M., et al. 2007, A\&A, 468, 979
\bibitem[]{} Marconi, A., \& Hunt, L. K. 2003, ApJ, 589, L21
\bibitem[]{} Menci, N., Fiore, F., Puccetti, S., \& Cavaliere, A. 2008, ApJ, 686, 219
\bibitem[]{} Mor, R., Netzer, H., \& Elitzur, M. 2009, ApJ, 705, 298
\bibitem[]{} Murray, N., Quataert, E., \& Thompson, T. A. 2005, ApJ, 618, 569
\bibitem[Nicastro(2000)]{2000ApJ...530L..65N} Nicastro, F.\ 2000, \apjl, 530, L65
\bibitem[]{} Plotkin, R. M., et al. 2010, AJ, 139, 390
\bibitem[]{} Polletta, M., et al.  2008, ApJ, 675, 960
\bibitem[]{} Raimundo, S. I., Fabian, A. C., Bauer, F. E., Alexander, D. M., Brandt, W. N., Luo, B., Vasudevan, R. V., \& Xue, Y. Q. 2010, MNRAS, in press
\bibitem[]{} Rowan-Robinson, M., Valtchanov, I. \& Nandra, K., 2009, MNRAS, 397, 1326
\bibitem[]{} Scoville, N., \& Norman, C. 1995, ApJ, 451, 510
\bibitem[]{} Shakura, N. I., \& Sunyaev, R. A. 1973, A\&A, 24, 337
\bibitem[]{} Shang, Z., et al. 2005, ApJ, 619, 41
\bibitem[]{} Shankar, F., Weinberg, D. H., \& Miralda-Escud\'{e}, J. 2009, ApJ, 690, 20
\bibitem[]{} Shin, M. S., Ostriker, J. P., \& Ciotti, L. 2010, ApJ, 711, 268
\bibitem[]{} Silk, J., \& Rees, M. J. 1998, A\&A, 331, L1
\bibitem[]{} Springel, V., DiMatteo, T., \& Hernquist, L. 2005, MNRAS, 361, 776
\bibitem[]{} Sunyaev, R. A., \& Titarchuk, L. G. 1985, A\&A, 143, 374
\bibitem[]{} Tremaine, S., et al. 2002, ApJ, 574, 740
\bibitem[]{} Treister, E., Krolik, J. H., \& Dullemond, C. 2008, ApJ, 679, 140
\bibitem[]{} Treister, E., Natarajan, P., Sanders, D., Urry, C. M., Schawinski, K., \& Kartaltepe, J. 2010, Science, 328, 600
\bibitem[]{} Ueda, Y., Akiyama, M., Ohta, K., \& Miyaji, T. 2003, ApJ, 598, 88
\bibitem[]{} Volonteri, M., Madau, P., Quataert, E., \& Rees, M. J. 2005, ApJ, 620, 69
\bibitem[]{} Vignali, C., Alexander, D. M., \& Comastri, A. 2006, MNRAS, 373, 321
\bibitem[]{} Wang, J., et al. 2009, ApJ, 697, L141
\bibitem[Zhang et al.(2009)]{2009ApJ...699..281Z} Zhang, W.~M., Soria, R., Zhang, S.~N., Swartz, D.~A., \& Liu, J.~F.\ 2009, \apj, 699, 281
\bibitem[Zhu et al.(2009)]{2009ApJ...700.1173Z} Zhu, L., Zhang, S.~N., \& Tang, S.\ 2009, \apj, 700, 1173




\end{thebibliography}
\end{document}